# Prediction of repurposed drugs for treating lung injury in COVID-19

Bing He[1], Lana Garmire[1*]

1. Department of Computational Medicine and Bioinformatics, Medical School, University of Michigan, Ann Arbor, 48105, USA

* To whom correspondence should be addressed.

Email address: lgarmire@med.umich.edu

# Abstract

Coronavirus disease (COVID-19) is an infectious disease discovered in 2019 and currently in outbreak across the world. Lung injury with severe respiratory failure is the leading cause of death in COVID-19, brought by severe acute respiratory syndrome coronavirus 2 (SARS-CoV-2). However, there still lacks efficient treatment for COVID-19 induced lung injury and acute respiratory failure. Inhibition of Angiotensin-converting enzyme 2 (ACE2) caused by spike protein of SARS-CoV-2 is the most plausible mechanism of lung injury in COVID-19. We propose two candidate drugs, COL-3 (a chemically modified tetracycline) and CGP-60474 (a cyclin-dependent kinase inhibitor), for treating lung injuries in COVID-19, based on their abilities to reverse the gene expression patterns in HCC515 cells treated with ACE2 inhibitor and in human COVID-19 patient lung tissues. Further bioinformatics analysis shows that twelve significantly enriched pathways (P-value <0.05) overlap between HCC515 cells treated with ACE2 inhibitor and human COVID-19 patient lung tissues, including signaling pathways known to be associated with lung injury such as TNF signaling, MAPK signaling and Chemokine signaling pathways. All these twelve pathways are targeted in COL-3 treated HCC515 cells, in which genes such as RHOA, RAC2, FAS, CDC42 have reduced expression. CGP-60474 shares eleven of twelve pathways with COL-3 with common target genes such as RHOA. It also uniquely targets genes related to lung injury, such as CALR and MMP14. In summary, this study shows that ACE2 inhibition is likely part of the mechanisms leading to lung injury in COVID-19, and that compounds such as COL-3 and CGP-60474 have the potential as repurposed drugs for its treatment.

# Background

Coronavirus disease (COVID-19) is an infectious disease discovered in 2019 and currently in outbreak across the world, resulting in more than 2.2 million infections and over 150 thousand of deaths by now. It is causing tens of thousands of new infections and thousands of mortalities every day. Patients with COVID-19 present with respiratory symptoms. Severe viral pneumonia related lung injury with acute respiratory failure is the main reason of COVID-19 related death[1]. However, there still lacks efficient treatment for COVID-19 induced lung injury and acute respiratory failure.

Coronaviruses (CoVs), are a large family of enveloped, positive-sense, single-stranded RNA viruses, which can be found in many vertebrates, such as birds, pigs and human, to cause various diseases. A novel CoV, termed severe acute respiratory synrdrome (SARS)-CoV-2, is the cause of COVID-19. Lung injury with acute respiratory failure was also the main reason of death in patients with SARS[2]. The spike protein of SARS-CoV-2 shares 79.5% sequence identity with SARS-CoV virus [3-5], which caused SARS pandemic in 2002 resulting in 774 deaths in 8096 confirmed patients in 29 countries [6]. SARS-CoV-2 uses angiotensin-converting enzyme 2 (ACE2) as the entry receptor and cellular serine protease TMPRSS2 for S protein for priming to allow fusion of viral and cellular membranes[7], similar to SARS-CoV [8, 9]. Since in SARS-CoV infection, spike protein of the SARS-CoV inhibits ACE2 to cause severe lung injury and acute respiratory failure[10, 11], it is highly likely that SARS-CoV-2 uses the same mechanism. Inhibition of ACE2 may be part of the pathogenic mechanism in SARS-CoV-2 induced lung injury and acute respiratory failure. Therefore, a drug repurposing pipeline aiming for reversing gene expression pattern due to ACE2 inhibition may be a candidate for treating lung injury in COVID-19.

Towards this goal, we performed drug repositioning analysis to identify drugs and compounds for treating SARS-CoV-2 induced lung injury. To explore the mechanisms of proposed drug treatment, we further investigated deregulated genes and pathways in both human lung cells treated with ACE2 inhibitor and human lung tissues from patients deceased from COVID-19. Our results revealed that lung injury related molecular mechanisms are shared between inhibition of ACE2 and infection of SARS-CoV-2. Moreover, our proposed drugs can target key genes in these mechanisms, and therefore may prevent lung injury in COVID-19.

# Methods

**Data Preparation**

RNA-seq data from human lung tissues from two COVID-19 deceased patients and age matched healthy lung tissues, as well as human lung A549 cells with or without H1N1 infection, were downloaded from Gene Expression Omnibus (GEO) database (GEO id: GSE147507), as reported by Melo et al. [12]. Level 5 LINCS L1000 data, a collection of gene expression profiles for thousands of perturbagens at a variety of time points, doses, and cell lines, were downloaded from GEO database (GEO id: GSE70138 and GSE92742). Gene expression profiles in lung cells were extracted from downloaded L1000 dataset. The extracted data include 37,366 treatments of 12,707 drugs in 13 lung cell lines at different time points and doses. Two lung cell lines, A549 and HCC515 were treated with 10 µM moexipril, homologue of ACE2 that inhibits ACE2 and ACE. Gene expression profiles were collected from A549 and HCC515 cells at 6 and 24 h after treatment. Upon moexipril treatment, ACE2 level decreased with time in HCC515 as expected, however increased in A549. This prompted us to focus the analysis using HCC515 line which showed the inhibition effect of moexipril. Differential expression of genes was measured by z-score[13].

**Gene and Pathway Analysis**

The RNA-seq data were analyzed using DESeq2. Differential gene expressions were identified by comparing between cases and controls (eg. COVID-19 lung tissue vs. the healthy lung tissue, or cells with H1N1 infection vs. those without H1N1 infection). Top 1000 differential expressed genes were selected by the absolute value of z-score. These genes were then used for pathways enrichment analysis using Database for Annotation, Visualization and Integrated Discovery (DAVID) v6.8[14]. Significant pathways (P-value <0.05) were compared between HCC515 cells with ACE2 inhibitor inhibition and lung tissues from COVID-19 deceased patients. A gene is called "consistent", if it shows changes in the same direction (increase or decrease) between treatment with ACE2 inhibitor and that with SARS-CoV-2 infection. The importance of pathways was ranked by the following score:

$$Score_{Pathway} = \sqrt{-\log \frac{Pvalue_{ACE2i} + Pvalue_{COVID19}}{2} n_{consistent}}$$

$Pvalue_{ACE2i}$ is the P-value from pathway enrichment analysis for the top 1000 differentially expressed genes in cells treated with ACE2 inhibitor. $Pvalue_{COVID19}$ is the P-value from pathway enrichment analysis for the top 1000 differentially expressed genes in lung tissue infected by SARS-CoV-2. $n_{consistent}$ is the number of consistent genes in that pathway.

The importance of genes was ranked by the following score:

$$Score_{Gene} = \sqrt{\frac{Zscore_{ACE2i} + Zscore_{COVID19}}{2} n_{pathway}}$$

$Zscore_{ACE2i}$ is the z-score of the gene in cells treated with ACE2 inhibitor. $Zscore_{COVID19}$ is the z-score of the gene in lung tissue infected by SARS-CoV-2. $n_{pathway}$ is the number of significant pathways this gene belongs to.

**Drug Repositioning Analysis**

Differential gene expression list was transformed to a gene rank list. An effective drug treatment is one that reverts the aberrant gene expression in disease back to the normal level in health DrInsight Package[15] was used for this purpose, and the outlier-sum (OS) based statistic was retrieved, which models the overall disease-drug connectivity by aggregating disease transcriptome changes with drug perturbation. The Kolmogorov–Smirnov (K-S) test was then applied on the OS statistic, to show the significance level of one drug treatment, relative to the background of all other drugs and compounds in the reference drug dataset. The reference drug dataset contains gene rank lists from 12,707 drug treatments in the LINCS L1000 data, as mentioned above. The false discovery rate (FDR) was used to adjust P-values from the K-S test, to avoid false significance due to multiple comparisons. FDR<0.05 was used as the threshold to select significant drug candidates for the disease.

## Results

**Feasibility test of the drug repositioning pipeline using influenza A (H1N1) infection data**

Our drug repositioning is based on the assumption that if a drug can reverse the abnormality of gene expression pattern in the disease, the drug should be able to treat the disease [16, 17]. Towards this we have implemented the computational framework as shown in Figure 1. We collected differential gene expression patterns in the disease and in cells with drug treatment. Then we searched reversible genes whose expression changes in drug treatment are opposite to these in disease to estimate the effect of a drug for the disease. We further compared effect of every drug to all other candidates to estimate the significance of a drug for treating the disease.

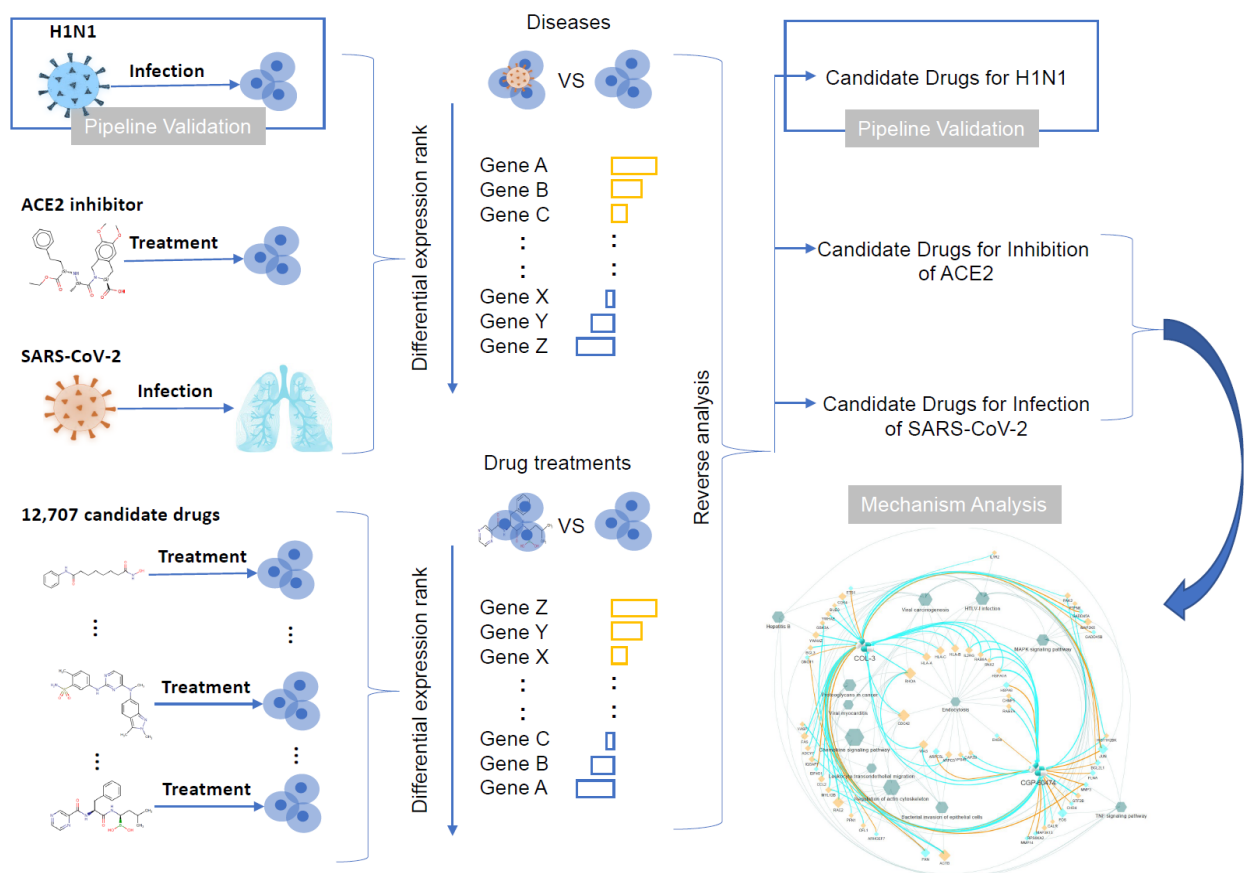

**Figure 1. Workflow of repurposing drugs for treating lung injury in COVID-19.** Input data include gene expression in A549 cells with H1N1 infection, HCC515 cells with ACE2 inhibitor (ACE2i), human lung tissues from COVID-19 deceased patients and cells with drug treatment, respectively. Reversing analysis is conducted to search for drugs which can reverse the gene expression changes upon the treatment. The candidate drug to is compared to all other drugs and compounds, in order to estimate its significance level at treating the disease. Candidate drugs for H1N1 are used for validation of the computational pipeline. Candidate drugs identified in both HCC515 cells treated with ACE2 inhibitor and in human lung tissues from COVID-19 deceased patients are used for downstream mechanism analysis.

As COVID-19 is an emerging disease with much unknown, we first demonstrate the feasibility the drug repositioning pipeline using infection of H1N1 virus, where much more research has been done and multiple drugs are approved by FDA. We computed the differentially expressed genes from RNA-seq data of A549 lung cells with or without H1N1 virus infection. We then identified the best candidates that could reverse the expression pattern of these differentially expressed genes, by analyzing 12,707 drugs and compounds from LINCS L1000 pharmacogenomics data [13]. The results show that CGP-60474 (FDR= $2.514×10^{-4}$), Sirolimus (FDR= $3.040×10^{-4}$), COL-3 (FDR= $9.452×10^{-4}$), PIK-75 (FDR= 0.002), and Wortmannin (FDR= 0.046) could significantly (FDR<0.05) reverse the gene expression in H1N1 infection in A549 lung cells (Table 1). Sirolimus, the second-best candidate by FDR, also known as rapamycin, is a potent immunosuppressant that acts by selectively blocking the transcriptional activation of cytokines thereby inhibiting cytokine production. It was previously shown clinically effective in H1N1 infected patients with severe pneumonia and acute respiratory failure[18], as adjuvant treatment with steroids. In summary, our drug repositioning pipeline has shown merit in discovering effective drugs, through the example of H1N1 infection.

**Repurposed drugs for treating lung injury in COVID-19**

To repurpose drugs for inhibition of ACE2, we conducted differential gene expression analysis in HCC515 and A549 lung cells with the inhibition of ACE2 by moexipril, from LINCS L1000 project[13]. using the similar approach as the H1N1 infection described above. Upon examination of ACE2 expression at different time points (6h and 24h), we opted to focus on HCC515 cells which have reduced ACE2 expression over treatment of moexipril, an ACE2 inhibitor. At 6 h after treatment of moexipril, narciclasine (FDR=0.006) and geldanamycin (FDR=0.006) could significantly reverse the gene expression changes due to ACE2 inhibitor (Table 1). At 24h post treatment of moexipril, the effect of CGP-60474 (FDR=$1.337×10^{-7}$), panobinostat (FDR=$2.443×10^{-05}$), trichostatin-a (FDR= $3.546×10^{-03}$) and COL-3 (FDR= 0.002) became significant (Table 1).

To further confirm if these effects shown from cell lines are physiologically relevant for human lung injury due to COVID-19, we analyzed the RNA-Seq data of human lung tissues from 2 COVID-19 deceased patients with age-matched normal lung tissues, as reported by Melo et al. [12]. A check on gene expression of individual markers for lung injury, advanced glycosylation end-product specific receptor (AGER), lipopolysaccharide binding protein (LBP) and secretoglobin family 1A member (SCGB1A1) [19], shows that they are up-regulated in HCC515 cell line treated with ACE2 inhibitor and human COVID-19 patient lung tissue (Figure 2). Whereas surfactant protein D (SFTPD), a gene encoding protein that innates immune response to protect the lungs against inhaled microorganisms and chemicals, is decreased. This indicates the similarity between ACE2 inhibition by moexipril in the cell line and lung injury from COVID-19. Next we extracted the differentially expressed genes in COVID-19 lung tissues vs. normal lungs, and used them as target genes to be reversed by the same drugs and compounds in the drug repositioning framework as shown in Figure 1. The results show that Sirolimus (FDR=0.003), COL-3 (FDR=0.003), CGP-60474 (FDR=0.003), Staurosporine (FDR=0.003) and Mitoxantrone (FDR=0.003) are significant in reversing the target genes' expression in the human lung tissues due to COVID-19 mentioned earlier (Table 1). Thus, together COL-3 and CGP-60474 show consistent effects for reversing gene expression changes in both HCC515 cell line treated with ACE2 inhibitor and human COVID-19 patient lung tissue (Table 1). Moreover, COL-3 and CGP-60474 both can reversely decrease the expression of marker genes for lung injury, AGER, LBP, SCGB1A1, and reversely increase SFTPD expression in HCC515 cell line pre-treated with ACE2 inhibitor moexipril. CGP-60474 (0.12 µM) appears to be more potent than COL-3 (2.5 µM). In conclusion, COL-3 and CGP-60474 show promises as potential purposeful drugs to treat lung injury in COVID-19.

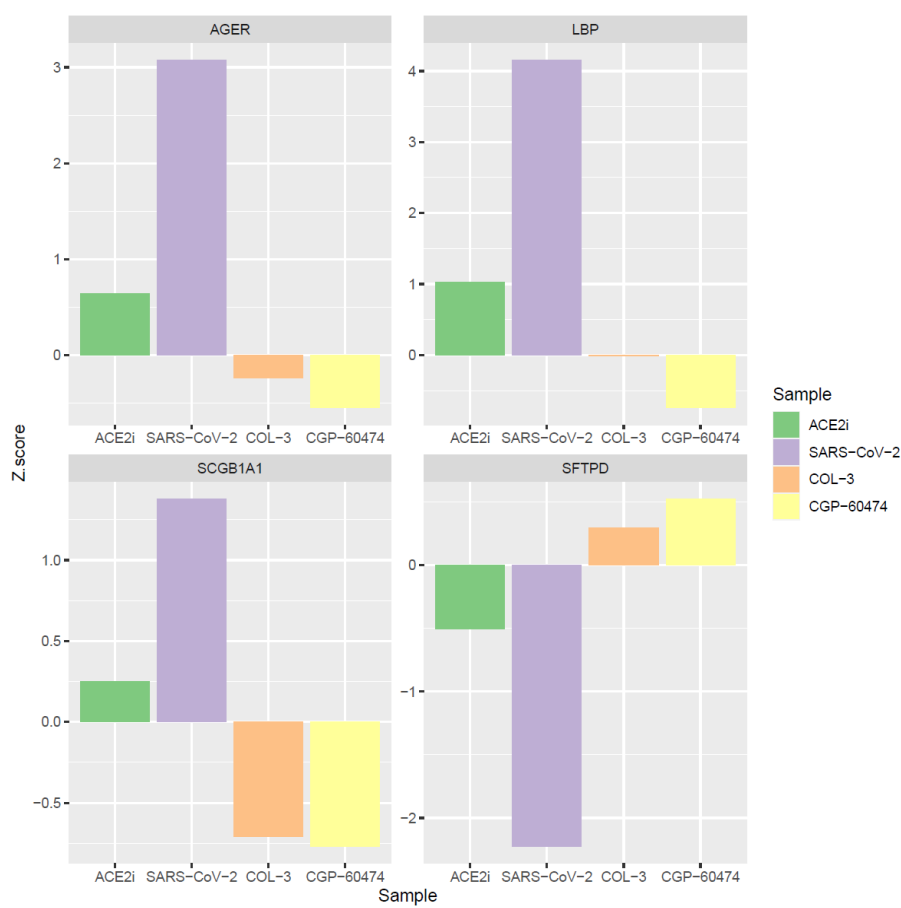

**Figure 2. COL-3 and CGP-60474 can reverse the expression of marker genes of lung injury.** Z-score: z score of differential expression of genes in the sample; ACE2i: HCC515 cells with ACE2 inhibitor inhibition;

SARS-CoV-2: human lung tissues from COVID-19 patients deceased from SARS-CoV-2 induced lung complications; COL-3: HCC515 cells treated with COL-3; CGP-60474: HCC515 cells treated with CGP-60474.

**Pathway comparison between inhibition of ACE2 and infection of SARS-CoV-2**

We performed pathway enrichment analysis with top 1000 deregulated genes in HCC515 cells with ACE2 inhibitor inhibition and human COVID-19 patient lung tissues, respectively. Twelve significantly enriched pathways (P-value <0.05) overlap between HCC515 cells with ACE2 inhibitor inhibition and human COVID-19 patient lung tissues (Figure 3, Table 2). As expected, multiple pathways involved in virus infection are enriched. Various signaling pathways, such as TNF signaling pathway, MAPK signaling pathway and Chemokine signaling pathway are also enriched, with well-known associations with lung injury[20-22]. Moreover, other pathways related to cancers (eg. "Viral carcinogenesis" and "Proteoglycans in cancer"), or cardiovascular diseases (eg. Viral myocarditis) also show up significantly enriched in the results (Figure 3, Table 2). Sixty-six genes in these overlapped pathways show consistent changes between ACE2 inhibited lung cell line and SARS-CoV-2 lung tissues (Table 2).

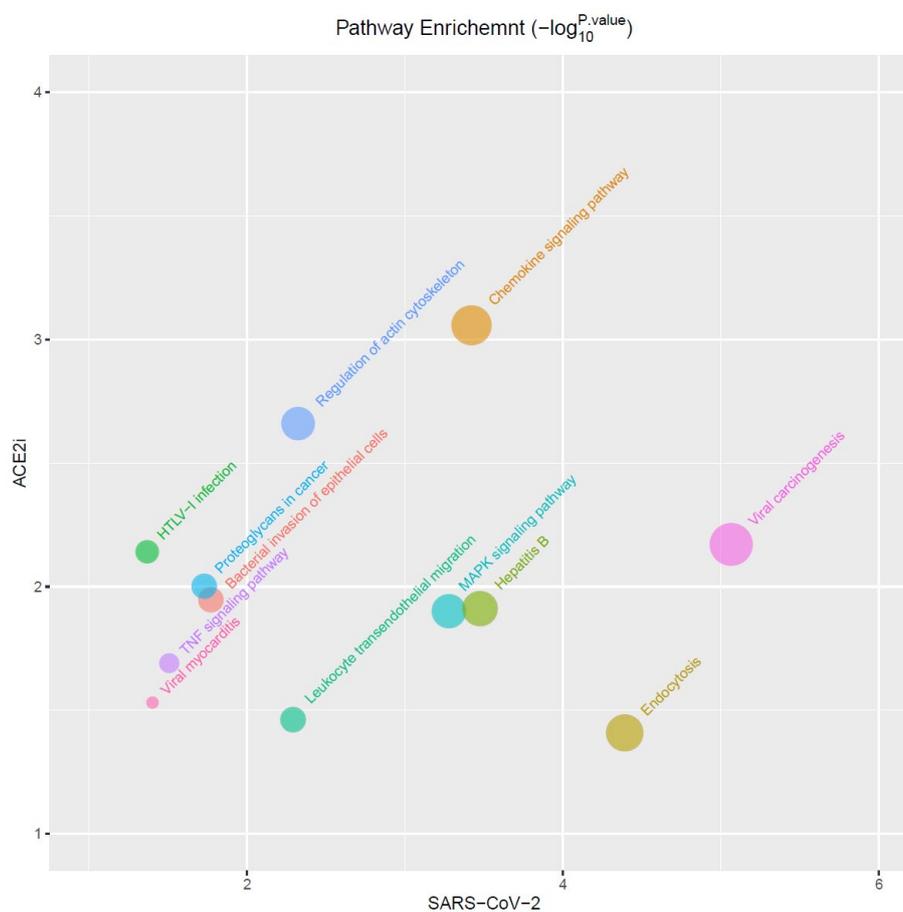

**Figure 3. The bubble plot of significantly enriched pathways in HCC515 cells with ACE2 inhibitor inhibition and human COVID-19 patient lung tissues.** X-axis and Y-axis show -log10 transformed P-values in human COVID-19 patient lung tissues (SARS-CoV-2) and HCC515 cells with ACE2 inhibitor inhibition (ACE2i), respectively. Size of the bubble shows the average value of -log10 transformed P-value in SARS-CoV-2 and ACE2i.

We further analyzed the genes and pathways associated with the two drugs COL-3 and CGP-60474, which show coherent effects in reversing the gene expression patterns in HCC515 cells with ACE2 inhibitor inhibition and human COVID-19 patient lung tissues (Figure 4). For COL-3, from the molecular point of view, it leads to decreased expression of many genes including RHOA, RAC2, FAS and CDC42 in lung cells, as part of the mechanisms to protect lung from injury (Figure 4). These genes are important plays in pathways such as Chemokine signaling pathway (for CCL2, ADCY7, GNG11, PXN, CDC42, RAC2, RHOA, WAS), TNF signaling pathway (for CCL2, MMP3, JUN, BCL3, FAS, MAP2K6) and MAPK signaling pathway (for HSPA1A, CDC42, RAC2, PAK2, FAS, MAP2K6, JUN, GADD45B, GADD45A). All twelve significantly enriched pathways in Figure 3 are also observed in COL-3 treatment. For CGP-60474, it shares thirteen gene targets with COL-3, including RHOA, WAS, HSPA1A, SNX2, RAB8A, IL2RG, MMP3, BCL2L1, JUN, HIST1H2BK, GNG11, IQGAP1 and MYL12B. It also has a unique set of target genes related to lung injury, such as CALR and MMP14 (Figure 4). It decreases the expression of CALR, a multifunctional protein that acts as a major Ca(2+)-binding (storage) protein in the lumen of the endoplasmic reticulum [23]. It also increases the expression of MMP14, a member of the matrix metalloproteinase (MMP) family with anti-inflammatory property. CGP-60474 treatment affects eleven out of twelve significantly enriched pathways in COL-3, but not Viral myocarditis pathway. More details on the molecular mechanisms of the target genes and pathways of these two drug candidates are discussed below.

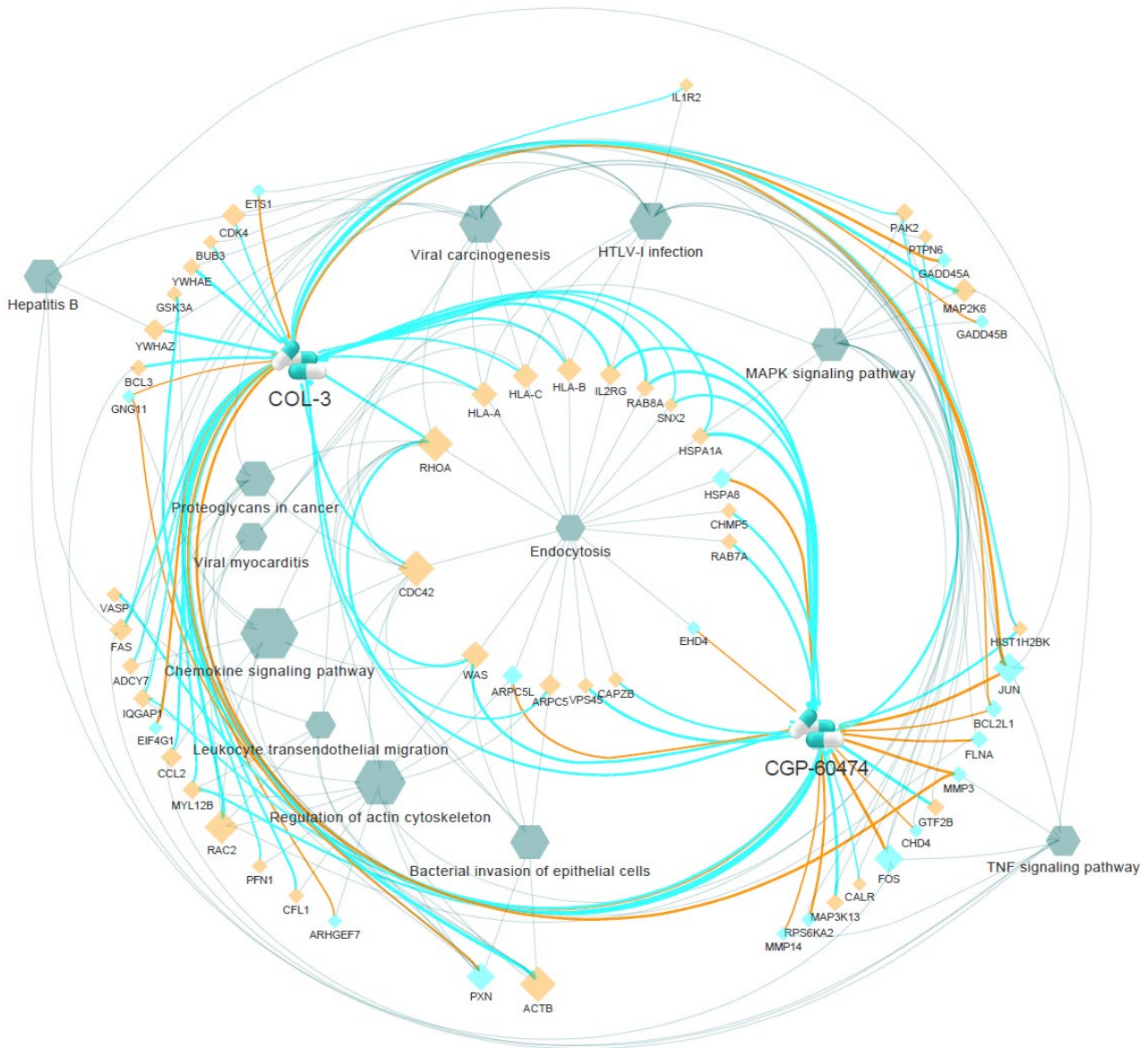

**Figure 4.** Target genes and pathways of COL-3 and CGP-60474 in treating lung injury in COVID-19. All pathways were significant enriched in both human COVID-19 patient lung tissues and HCC515 cells with ACE2 inhibitor inhibition. The abnormal gene expression patterns in these pathways were reversed by COL-3 and/or CGP-60474. Blue diamond: Down-regulated gene in disease; Orange diamond: Up-regulated gene in disease; Hexagon: Pathway; Blue line: Drug decreases gene expression; Orange line: drug increases gene expression; Blue/Orange line width corresponds to the ability to change gene expression; Dark green line: Interaction between gene and pathway; Diamond size: Importance of gene in the disease; Hexagon size: Importance of pathway in the disease.

## Discussion

The inhibition of ACE2 promotes lung injury via the renin–angiotensin system (RAS)[24]. In pulmonary RAS, ACE2 converts angiotensin II (Ang II), an octapeptide hormone, to Ang-(1-

7), an heptapeptide hormone (Figure 5). Ang II triggers pulmonary inflammation and activates TNF signaling pathway and MAPK signaling pathway to promote lung injury[25, 26]. On the other hand, Ang-(1–7) inhibits inflammation protects lung from injury[27], by inhibiting MAPK signaling pathway[28], lowering cytokine release[29] and downregulating RHOA/ROCK pathway[30]. Thus, inhibition of ACE2 will increase Ang II level, decrease Ang-(1–7), and deregulate various downstream pathways, such as TNF and MAPK signaling pathways to promote lung injury (Figure 5). Our pathway analysis on HCC515 lung cell line confirmed that inhibition of ACE2 by moexipril can deregulate TNF signaling, MAPK signaling and cytokine signaling pathways. We further showed that these pathways are also deregulated in human lung tissues from COVID-19 deceased patients (Table 2). Moreover, inhibition of ACE2 induced similar expression patterns of lung injury markers to that in human lung tissues from COVID-19 deceased patients (Figure 2). All these evidences suggest that inhibition of ACE2 may indeed be part of the molecular mechanisms of lung injury in COVID-19. Moreover, other pathways related to cancers (eg. "Viral carcinogenesis" and "Proteoglycans in cancer"), or cardiovascular diseases (eg. viral myocarditis) also show up significantly enriched in the results (Table 2). These results may help to explain the increased risks of fatality among COVID-19 patients with underline conditions (cancers, heart diseases)[31, 32]. Additionally, myocarditis has been clinical observed in a patient with COVID-19[33], showing direct link between the two conditions.

Our drug repositioning analysis suggested five possible drugs based on RNA-Seq data from patients deceased from COVID-19. Among them, Sirolimus has started clinical trial for treating patients with COVID-19 pneumonia (Clinical trial: NCT04341675). Two other drugs (or compounds) COL-3 and CGP-60474 also have additional evidence of effectiveness from the L1000 data of lung HCC515 cell line treated with ACE2 inhibitor moexipril. Moreover, both COL-3 and CGP-60474 could reverse the expression patterns of lung injury markers in HCC515 cells with ACE2 inhibitor inhibition and human COVID-19 patient lung tissues (Figure 2). All these "phenotypic" evidences suggest that COL-3 and CGP-60474 may be effective in treating lung injury in COVID-19 (Figure 5). Therefore, we further analyzed the target genes and pathways of these two drugs in treating lung injury in COVID-19.

COL-3, also known as incyclinide or CMT-3, is a chemically modified tetracycline. It reversed the expression patterns of many lung injury related genes and pathways, such as RHOA, RAC2 and FAS in Chemokine signaling pathway, TNF signaling pathway and MAPK signaling pathway (Figure 4). RHOA, also known as ras homolog family member A, is a member of the Rho family of small GTPases. The activation of RHOA is crucial for lung injury[34]. Inhibition of RHOA is a promising approach to acute lung injury treatment[35]. RAC2, also known as Ras-related C3 botulinum toxin substrate 2, is a member of the Ras superfamily of small guanosine triphosphate (GTP)-metabolizing proteins. Rac2 plays an important role in inflammation-mediated lung injury[36, 37]. FAS, also known as Fas cell surface death receptor, is a member of the TNF-receptor superfamily. FAS activation is essential in inducing acute lung injury[38]. Small interfering RNA targeting Fas reduced lung injury in mice[39]. Previous results from many pre-clinical animal models have supported the role of COL-3 in reducing lung injury and improves survival of experimented animals. For example, COL-3 prevented lung injury and acute respiratory distress syndrome (ARDS) in clinically applicable porcine model[40-46]. It also improved acute respiratory distress syndrome (ARDS) survival in ovine model [47]. Given all the evidence, COL-3 will be an attractive candidate of clinical trial for treating severe viral pneumonia related lung injury with respiratory failure in COVID-19 (Figure 5).

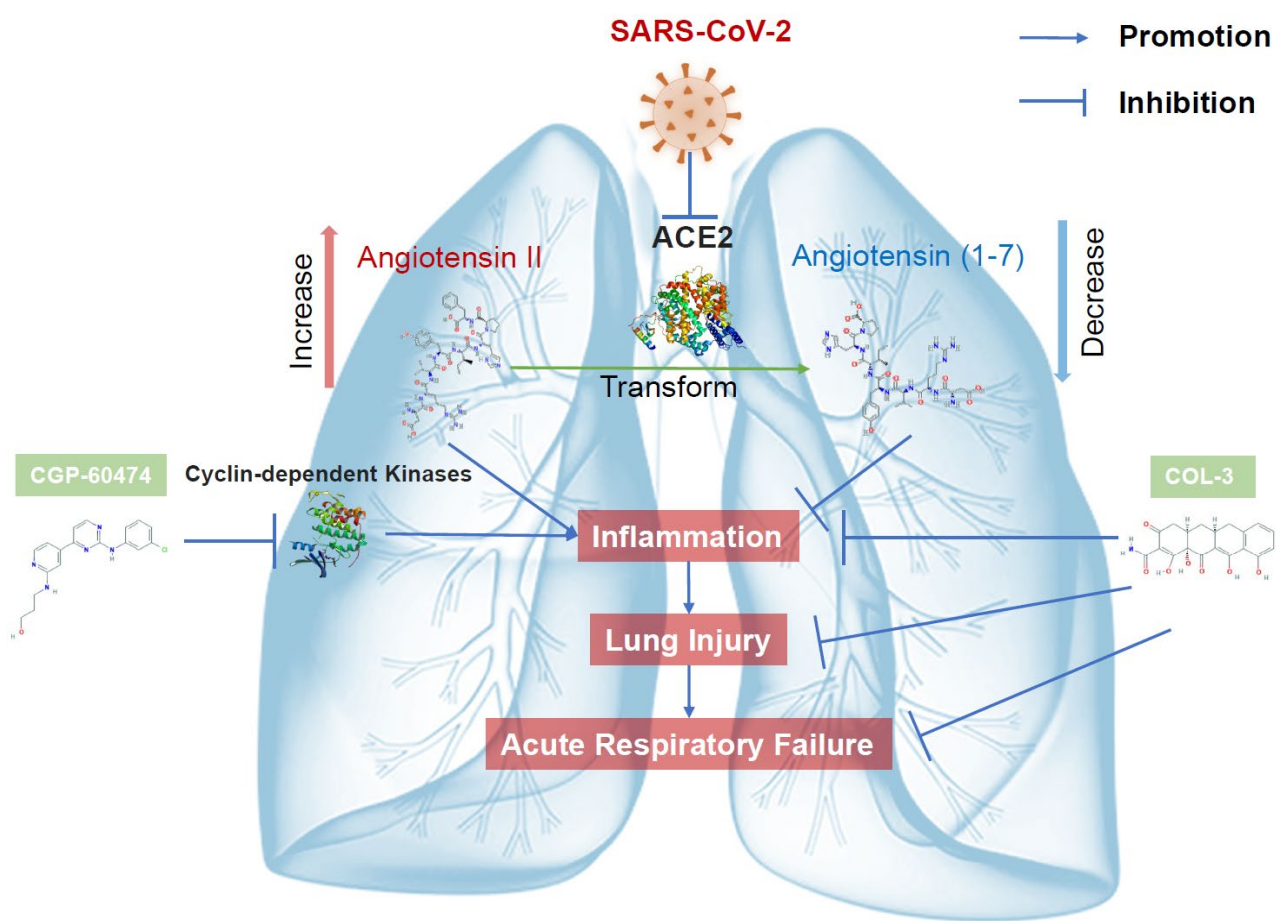

**Figure 5.** Proposed mechanisms of lung injury in COVID-19 through ACE2 and the therapeutic effects of COL-

3 and CGP-60474.

CGP-60474, on the other hand, is an inhibitor of cyclin-dependent kinase (Figure 5). CGP-60474 not only shared target genes with COL-3, such as RHOA, WAS, HSPA1A, SNX2, RAB8A, IL2RG, MMP3, BCL2L1, JUN, HIST1H2BK, GNG11, IQGAP1 and MYL12B, but also have unique target genes that related to lung injury, like CALR and MMP14 (Figure 4). Blocking CALR activity attenuated murine acute lung injury by inducing polarization of M2 subtype macrophages, which are anti-inflammatory[48]. MMP14 was shown to trigger the anti-inflammatory proteolytic cascade to prevent lung injury in mice[49]. Interestingly, so far only a few studies have reported some biological functions of CGP-60474 [50, 51] [52]. One drug reposition study using L1000 data also pointed to CGP-60474 as the most potent drug on the anti-inflammatory effects [52]. The authors then experimentally showed that CGP-60474 alleviated tumor necrosis factor-α (TNF-α) and interleukin-6 (IL-6) levels in activated macrophages, downregulated the NF-κB activity, and reduced the mortality rate in Lipopolysaccharide (LPS) induced endotoxemia mice. Another in silico drug prediction study suggested that CGP-60474 could target multiple cancers, though no experiments were conducted [50]. Although cyclin-dependent kinase inhibition by another drug Seliciclib reduced lung damage in a mouse model of ventilator-induced lung injury[51], further *in vivo* investigation of CGP-60474 are needed to test its role in treating lung injury.

## Conclusions

In summary, we propose two candidate drugs, COL-3 and CGP-60474, which can reverse the gene expression patterns in COVID-19 lung injury and lung cell line with ACE2 being inhibited. We further analyzed potential molecular and biological mechanisms of lung injury in COVID-19. The work will hopefully gain the interest of the biomedical and clinical community for further validations in vivo for both candidate drugs, and even possibly clinical trials on COL-3 to save lives from severe respiratory failure in COVID-19.

## List of abbreviations

COVID-19: Coronavirus disease

SARS-CoV-2: Severe acute respiratory syndrome coronavirus 2

ACE2: Angiotensin-converting enzyme 2

AGER: Advanced glycosylation end-product specific receptor

LBP: Lipopolysaccharide binding protein

SCGB1A1: Secretoglobin family 1A member

SFTPD: Surfactant protein D

RAS: Renin–angiotensin system

Ang II: Angiotensin II

Ang-(1-7): Angiotensin (1-7)

ARDS: Acute respiratory distress syndrome

ACE2i: Inhibition of ACE2

NS: Not significant

NA: Not available

## Declarations

### Competing interests

The authors declare no competing financial interests.

### Code Availability

All the codes and data are available at:

https://github.com/lanagarmire/COVID19-Drugs-LungInjury

### Funding

This research was supported by grants K01ES025434 awarded by NIEHS through funds provided by the trans-NIH Big Data to Knowledge (BD2K) initiative (www.bd2k.nih.gov), R01 LM012373 and R01 LM12907 awarded by NLM, and R01 HD084633 awarded by NICHD to L.X. Garmire.

### Authors' contributions

L.X. Garmire designed the project. B.He collected data and performed analysis. B.He and L.X. Garmire prepared tables, figures and wrote the manuscript.

**Table 1.** Significant candidate drugs for treating infection of H1N1, inhibition of ACE2 and infection of SARS-CoV-2, respectively.

| Drug | FDR value | | | |
|---|---|---|---|---|
| | H1N1 Infection | ACE2i | | SARS-CoV-2 Infection |
| | A549 cell | HCC515 cell | HCC515 cell | Human lung tissue |
| | 9h | 6h | 24h | NA |
| Sirolimus | $3.040 \times 10^{-4}$ | NS | NS | 0.003 |
| COL-3 | $9.452 \times 10^{-4}$ | NS | 0.002 | 0.003 |
| Geldanamycin | 0.001 | 0.006 | NS | NS |
| CGP-60474 | $2.514 \times 10^{-4}$ | NS | $1.337 \times 10^{-7}$ | 0.003 |
| Staurosporine | NS | NS | NS | 0.003 |
| Mitoxantrone | NS | NS | NS | 0.003 |
| Trichostatin-a | NS | NS | 0.004 | NS |
| Panobinostat | NS | NS | $2.443 \times 10^{-5}$ | NS |
| Narciclasine | NS | 0.006 | NS | NS |
| PIK-75 | 0.002 | NS | NS | NS |
| Wortmannin | 0.046 | NS | NS | NS |

NS: Not Significant.

NA: Not Available.

ACE2i: Inhibition of ACE2

**Table 2.** Pathway comparison between HCC515 cells with ACE2 inhibitor inhibition and human COVID-19 patient lung tissues.

| Pathway Name | P-value SARS-CoV-2 Human lung tissue | P-value ACE2i HCC515 cell | Consistent genes |
|---|---|---|---|
| Viral carcinogenesis | $8.610 \times 10^{-06}$ | $6.744 \times 10^{-03}$ | YWHAZ, PXN, CDC42, HIST1H2BK, RHOA, CHD4, TP53, HLA-A, HLA-C, HLA-B, CDK4, YWHAE, GTF2B, JUN |
| Endocytosis | $4.068 \times 10^{-05}$ | $3.902 \times 10^{-02}$ | RAB7A, CHMP5, SNX2, HSPA1A, ARPC5, CAPZB, CDC42, RHOA, IL2RG, HSPA8, EHD4, RAB8A, VPS45, HLA-A, HLA-C, HLA-B, WAS, ARPC5L, ARF3 |
| Hepatitis B | $3.354 \times 10^{-04}$ | $1.227 \times 10^{-02}$ | YWHAZ, TP53, RAF1, CDK4, STAT6, FOS, JUN, FAS |
| Chemokine signaling pathway | $3.797 \times 10^{-04}$ | $8.760 \times 10^{-04}$ | CCL2, ADCY7, GNG11, PXN, CDC42, RAC2, RHOA, RAF1, WAS, GSK3A, GNB1 |
| MAPK signaling pathway | $5.283 \times 10^{-04}$ | $1.257 \times 10^{-02}$ | HSPA1A, FOS, CDC42, RAC2, PAK2, FAS, MAP2K6, HSPA8, TP53, NR4A1, RAF1, FLNA, RPS6KA2, JUN, GADD45B, GADD45A, MAP3K13 |
| Regulation of actin cytoskeleton | $4.760 \times 10^{-03}$ | $2.189 \times 10^{-03}$ | ARPC5, PXN, IQGAP1, CDC42, PFN1, RAC2, PAK2, RHOA, ACTB, ARHGEF7, RAF1, MYL12B, WAS, ARPC5L, CFL1 |
| Leukocyte transendothelial migration | $5.122 \times 10^{-03}$ | $3.452 \times 10^{-02}$ | ACTB, MYL12B, PXN, VASP, CDC42, RAC2, RHOA |
| Bacterial invasion of epithelial cells | $1.697 \times 10^{-02}$ | $1.130 \times 10^{-02}$ | ACTB, CDC42, ARPC5L, RHOA, ARPC5, WAS, PXN |
| Proteoglycans in cancer | $1.870 \times 10^{-02}$ | $9.963 \times 10^{-03}$ | ACTB, PTPN6, TP53, RAF1, IQGAP1, PXN, FLNA, CDC42, RHOA, FAS |
| TNF signaling pathway | $3.117 \times 10^{-02}$ | $2.039 \times 10^{-02}$ | CFLAR, CCL2, MMP14, MMP3, FOS, JUN, BCL3, FAS, MAP2K6 |
| Viral myocarditis | $3.976 \times 10^{-02}$ | $2.943 \times 10^{-02}$ | ACTB, EIF4G1, RAC2, HLA-A, HLA-C, HLA-B |
| HTLV-I infection | $4.296 \times 10^{-02}$ | $7.225 \times 10^{-03}$ | IL1R2, ADCY7, BCL2L1, CALR, FOS, IL2RG, BUB3, EGR1, TP53, HLA-A, HLA-C, HLA-B, CDK4, ETS1, JUN |

ACE2i: Inhibition of ACE2